\documentclass[twoside]{dis08}
\usepackage[latin1]{inputenc}
\usepackage[dvips]{graphicx,epsfig,color}
\usepackage{wrapfig,rotating}
\usepackage{amssymb,amsmath,array}

\newcommand{\be}{\begin{equation}}
\newcommand{\ee}{\end{equation}}

\newcommand{\bl}{\pmb{l}}
\newcommand{\bk}{\pmb{k}}

\newcommand{\bea}{\begin{eqnarray}}
\newcommand{\eea}{\end{eqnarray}}
\newcommand{\cM}{{\cal M}}

\newcommand{\beq}{\begin{equation}}
\newcommand{\eeq}{\end{equation}}
\newcommand{\beqar}[1]{\begin{eqnarray}\label{#1}}
\newcommand{\eeqar}{\end{eqnarray}}

\newcommand{\beeq}{\begin{eqnarray}}
\newcommand{\eeeq}{\end{eqnarray}}

\begin{document}
\title{Searching for odderon in exclusive vector meson hadroproduction}


\author{Leszek Motyka$^{1,2}$
%
\thanks{The support of the DFG grant no.\ SFB676 is gratefully acknowledged.}
%
\vspace{.3cm}\\
%
1- II Institute for  Theoretical Physics, University of Hamburg,\\ 
Luruper Chaussee 149, D-22761, Germany \\[1mm]
2- Institute of Physics, Jagellonian University \\
Reymonta 4, 30-059 Krak\'{o}w, Poland
%
}


\maketitle

\begin{abstract}
In this talk \cite{url} estimates are presented of the odderon 
contributions to the exclusive $J/\psi$ and $\Upsilon$ production 
cross-sections at the Tevatron and the Large Hadron Collider. 
The obtained cross-sections are compared to cross-sections of 
the dominant background subprocesses mediated by the photon exchange. 
Possible experimental cuts are proposed that reduce the photon background.  
\end{abstract}

\section{Introduction}

Color neutral gluonic exchanges in high energy hadron scattering are 
naturally classified according to their $C$-parity. The $C$-even 
component is usually called the Pomeron. It gives the dominant contribution 
to vast majority of the measured cross-sections at high energies.
The $C$-odd partner of the Pomeron is the odderon. In contrast to the
Pomeron, the effects of the odderon are so weak that a compelling 
experimental evidence for the non-vanishing odderon contribution has not
been found yet. On the other hand, existence of the odderon exchange is 
guaranteed within perturbative QCD. 
In the lowest non-trivial order, the QCD odderon  
is made of three gluons with the symmetric color structure. Beyond the
leading order, the odderon amplitude can be described in the leading
logarithmic $\ln s$ approximation (where $\sqrt{s}$ is the collisions energy)
as a solution of the BKP evolution equation, a generalization of the famous 
BFKL equation for the Pomeron to a larger number of gluons. 
Thus, a successful measurement of the QCD odderon exchange should provide 
some important insight into the high energy evolution of multi-gluon 
amplitudes in QCD.

The main problem in odderon searches is the large background from the Pomeron
which, if present, prohibits a measurement of the odderon contribution.
In order to avoid this problem it is necessary to focus on processes 
in which, due to $C$-parity conservation, the Pomeron contribution vanishes.
This condition is fulfilled in exclusive production of mesons 
with definite $C$-parity, at high energies. A number of such measurements
was proposed and performed at HERA, unfortunately with negative results.

At hadron colliders, the Tevatron and the Large Hadron Collider (LHC) 
one expects to achieve an enhanced sensitivity to the odderon mediated 
processes, because of the strong coupling of proton projectiles to gluons. 
One of the simplest processes that should be sensitive to the odderon 
exchange in hadron collisions is the exclusive heavy vector meson 
production~\cite{first}, $\,p\bar p \to p \bar p + V\,$ or $\,p p \to p p + V\,$ 
where $V=J/\psi,\Upsilon$. 
In this talk~\cite{url} an attempt~\cite{paper} to estimate the 
corresponding cross-sections for the Tevatron and the LHC
is described and prospects for the measurements are discussed.

\begin{figure}[h]
\begin{center}
\begin{tabular}{llll}
\epsfxsize=0.2\columnwidth \epsffile{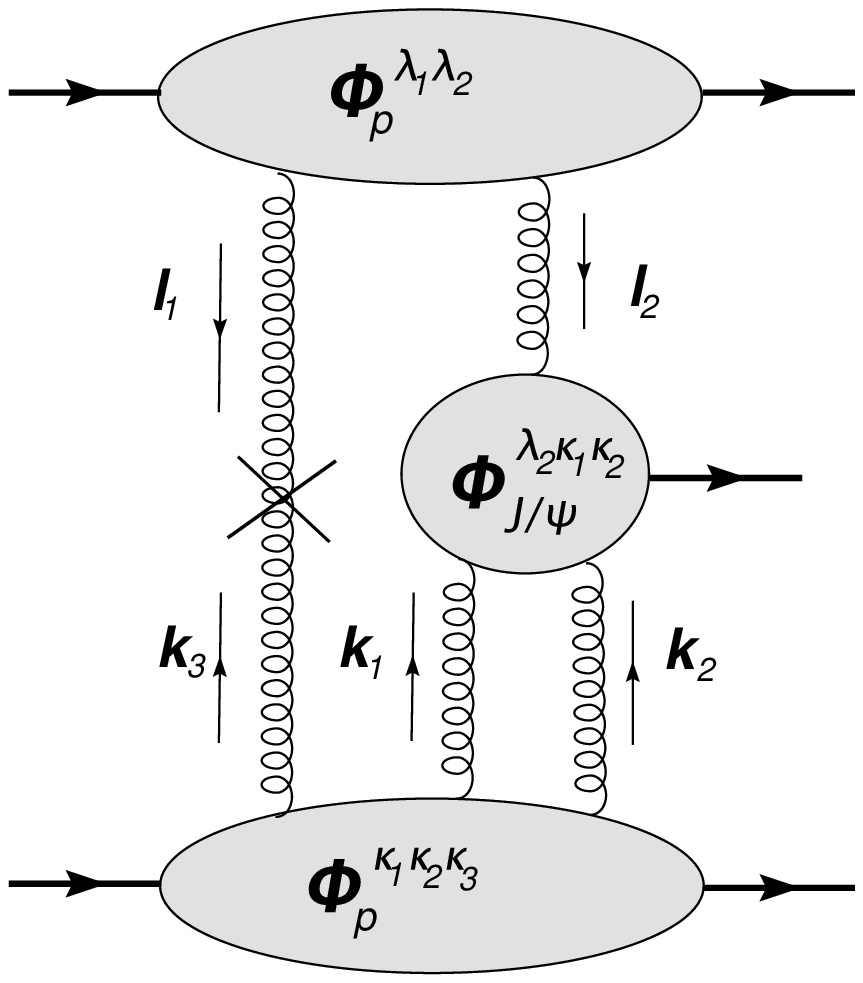} &
\epsfxsize= 0.2\columnwidth \epsffile{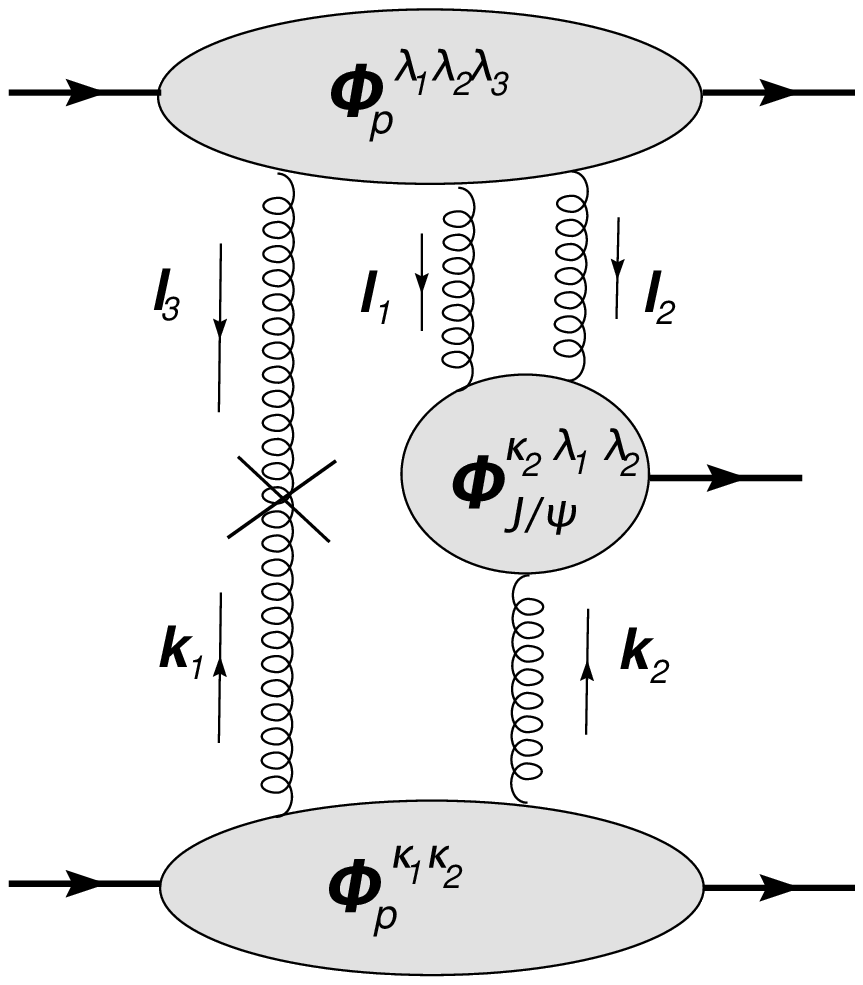} &
\epsfxsize=0.2\columnwidth \epsffile{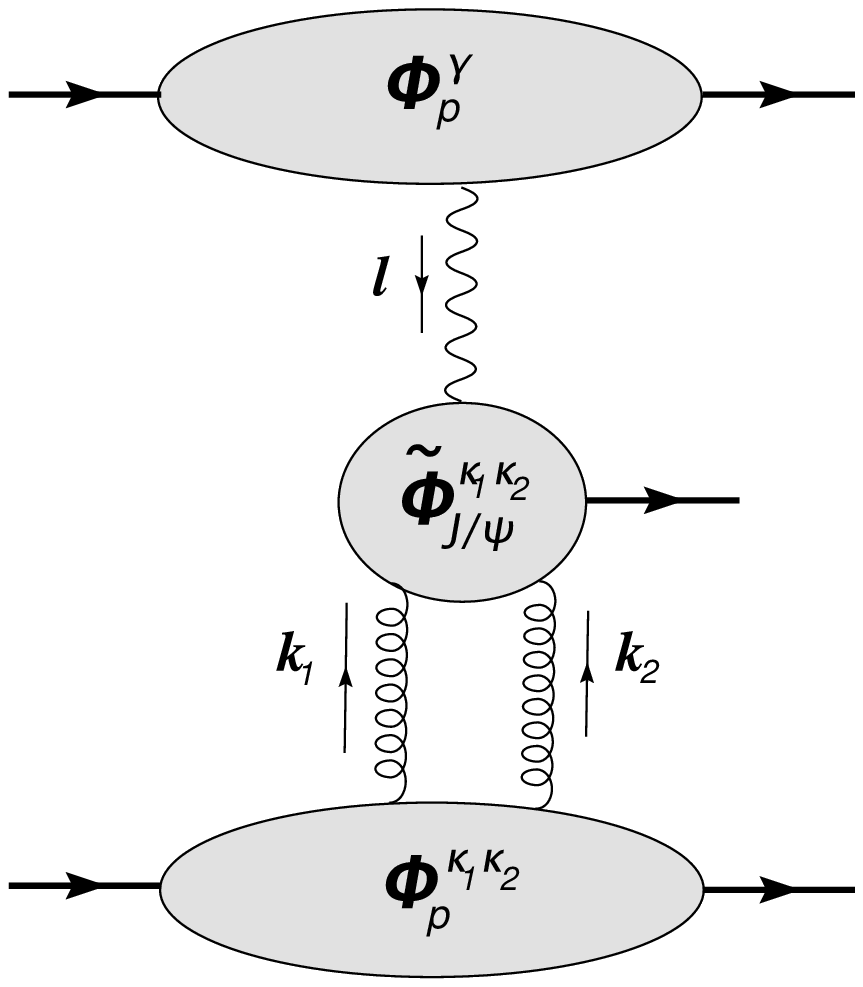}&
\epsfxsize= 0.2\columnwidth \epsffile{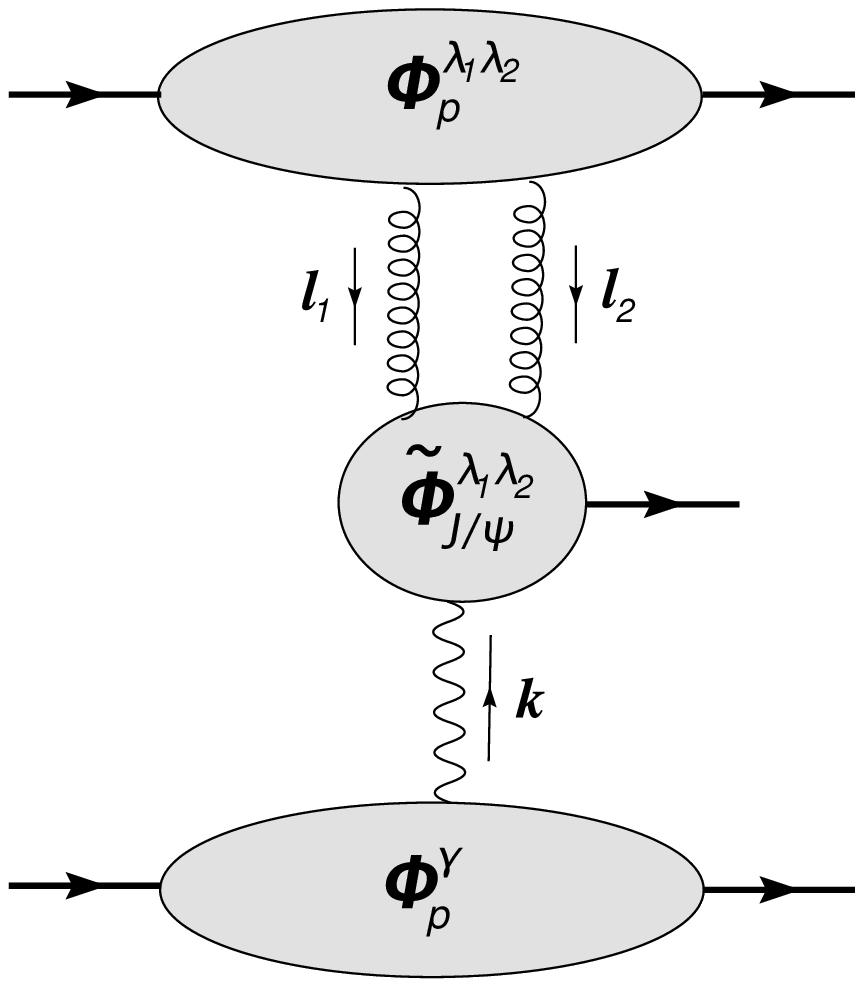} 
\\
a) & b) & c) & d) \\
\end{tabular}
\end{center}
\caption{The lowest order diagrams defining the 
Pomeron--odderon and Pomeron--photon fusion amplitudes of
the vector meson production 
a) ${\cal M}_{P\,O}$, 
b) ${\cal M}_{O\,P}$,
c) ${\cal M}_{\gamma\, P}$ and
d)  ${\cal M}_{P\, \gamma}$.}
\end{figure}

\section{Formalism}

There are two main components of the exclusive vector meson hadroproduction 
amplitudes: one coming from the odderon-Pomeron fusion (Fig.~1a,~1b) and the other coming from the photon-Pomeron fusion 
(Fig.~1c,~1d). We estimated both contributions within the $k_T$ 
factorization approach, at the lowest order. In this approximation, the
integrations over longitudinal components of loop momenta can be performed
exactly, and the non-trivial process dependent features are 
encoded in the impact factors $\Phi (\ldots)$, that describe couplings
of gluons to the scattering particles and depend only on the transverse
momenta of the gluons, $\bk_i$, $\bl_i$.
 As an example, we give here the amplitude of the diagram shown 
in Fig.~1a,
\bea
\label{impactA} 
&\cM_{P\;O} \,=\, -is\;\frac{2\cdot3}{2!\,3!}\,\frac{4}{(2\pi)^8}\,\int \frac{d^2 \bl_1}{\bl_1^2}\,\frac{d^2 \bl_2}{\bl_2^2}\ \delta^2(\bl_1 + \bl_2 - \bl)\,\frac{d^2 \bk_1}{\bk_1^2}\,\frac{d^2 \bk_2}{\bk_2^2}\,\frac{d^2 {\bk}_3}{\bk_3^2}\ \delta^2(\bk_1+\bk_2+\bk_3-\bk) &
\nonumber \\
&
\times \, \delta^2(\bk_3+\bl_1)\,\bk_3^2\;\delta^{\lambda_1 \kappa_3}
\cdot\Phi^{\lambda_1 \lambda_2}_P(\bl_1,\bl_2)\cdot\Phi^{\kappa_1 \kappa_2 \kappa_3}_O(\bk_1,\bk_2,\bk_3)\cdot
\Phi_{J/\psi}^{\lambda_2 \kappa_1 \kappa_2}(\bl_2, \bk_1,  \bk_2)\,
. &
\eea
Here  $\bk_i$, $\bl_i$ are gluon transverse momenta and $\lambda_i$,
$\kappa_i$ are gluon color indices.
$\Phi^{\lambda_1 \lambda_2}_P$ and  
$\Phi^{\kappa_1 \kappa_2 \kappa_3}_O$
denote the  impact factors of the  proton, scattered via  
the Pomeron and odderon exchange respectively.  
Both the impact factors are obtained in the 
Fukugita-Kwieci\'nski model of the proton, see \cite{paper} for details.
 The effective production vertex of the $J/\psi$~meson is denoted
$\Phi_{J/\psi}^{\lambda_2 \kappa_1 \kappa_2}$. 
It results from the perturbatively calculable fusion of three gluons into
$J/\psi$.
In order to keep the notation of momenta $\bl_i$ and $\bk_j$  most symmetric, 
we introduced an additional, artificial vertex (denoted by the cross 
in Fig.~1) $\delta^2(\bk_3+\bl_1)\,\bk_3^2\;\delta^{\lambda_1 \kappa_3}$
connecting the spectator gluons $(\bl_1,\lambda_1)$ and $(\bk_3,\kappa_3)$.
The ratio $\frac{2\cdot3}{2!\,3!}=\frac{1}{2}$ is a combinatorial factor. 
The factors $\frac{1}{2!}$ and $\frac{1}{3!}$ correct 
the  over-counting of diagrams introduced by factorization in the scattering 
amplitudes of the impact factor with Pomeron and odderon exchanges, 
respectively. The factor $2\cdot3=6$ accounts for all possibilities to
build the spectator gluon from the momenta $\bl_i$ and $\bk_j$.

\begin{table}[t]
\begin{center}%
\label{tab2}
\begin{tabular}[c]{|c|c|c|c|c|}\hline\hline
$d\sigma^{\mathrm{corr}}/dy$ & 
\multicolumn{2}{c|}{$ J/\psi$} &
\multicolumn{2}{c|}{$\Upsilon$} \\ \cline{2-5}
 &  odderon & photon & odderon & photon \\ \hline
Tevatron &
0.3--1.3--5~nb & 
0.8--5--9~nb &  
0.7--4--15~pb & 
0.8--5--9~pb  \\
LHC      & 
0.3--0.9--4~nb & 
2.4--15--27~nb  & 
1.7--5--21~pb & 
5--31--55~pb\\\hline \hline
\end{tabular}
\end{center}
\caption{ Estimates of cross-sections $d\sigma^{\mathrm{corr}} /dy|_{y=0}$ 
given for the exclusive $J/\psi$ and $\Upsilon$ production in $pp$ and 
$p\bar p$ collisions by the odderon--Pomeron fusion and the photon-Pomeron
fusion for the pessimistic--central--optimistic scenarios.}
\end{table}

\section{Results}

Diagrams shown in Fig.~1 give the lowest order amplitudes. A more realistic
estimate of the production amplitudes requires taking into account the QCD
evolution of the Pomeron. We represent this effect using a phenomenological enhancement factor $E(s,m_V)$, with $V=J/\psi,\Upsilon$. Besides that, it
is necessary to take into account corrections coming from multiple scatterings,
that may destroy the exclusive character of the process. Those effects will be expressed as a gap survival factor $S_{\mathrm{gap}}^2$.
An important model parameter that controls the magnitude of the proton
impact factors is an effective strong coupling constant, $\bar \alpha_s$. 
This parameter enters the Pomeron--odderon fusion 
cross-section in the fifth power. Thus, a cross-section, that takes into 
account necessary phenomenological improvements may be written as
$
\left.
{d \sigma^{\mathrm{corr}} / dy} 
\right|_{y=0}
\, = \, 
\bar\alpha_s^5\, S_{\mathrm{gap}}^2\, E(s,m_V) \,{d \sigma / dy},
\label{master}
$
where ${d \sigma/ dy}$ is the lowest order cross-section evaluated 
at $\bar\alpha_s=1$. 
We approximate the effects of QCD evolution of the Pomeron amplitude 
by an exponential enhancement factor $\exp(\lambda \, \Delta y)$ where  
$\Delta y$ is the rapidity evolution length of the QCD Pomeron. 
Thus, for the central production one obtains 
$E(s,m_V) = (x_0 \sqrt{s}/m_V)^{2\lambda}$, and the initial $x$-value 
for the gluon evolution is assumed to be $x_0 = 0.1$.
Following results from HERA, we take the effective Pomeron intercept 
$\lambda = 0.2$ ($\lambda = 0.35$) for the $J/\psi$ ($\Upsilon$) 
production.

The estimate of uncertainties introduced by $\bar\alpha_s$ and 
$S_{\mathrm{gap}}^2$ is carried out together. By doing so, we follow the
assumptions made in existing determinations of $\bar\alpha_s$.  
For instance, a low value of $\bar\alpha_s \simeq 0.3$ was obtained from an
analysis of the elastic $pp$ and $p\bar p$ scattering data 
in which $S_{\mathrm{gap}}^2=1$ was taken. 
Analyzes of inclusive cross-sections and the exclusive vector meson photoproduction yielded $\bar\alpha_s \simeq 0.7\,-\,1$.
Thus, we use  $S_{\mathrm{gap}}^2=1$ in our calculation if the 
low value of $\bar\alpha_s=0.3$ is taken. 
This combination $S_{\mathrm{gap}}^2=1$ and $\bar\alpha_s=0.3$ gives low 
cross-sections and it is called the {\em pessimistic scenario}.
In the {\em optimistic scenario} we use a $\bar\alpha_s = 1$, combined with 
the gap survival factors  obtained in the Durham two-channel eikonal model: 
$S_{\mathrm{gap}}^2 =0.05$ for the exclusive production at the Tevatron 
and $S_{\mathrm{gap}}^2=0.03$ for the LHC. The best 
estimates should follow from the {\em central scenario} defined by 
$\bar\alpha_s=0.75$, $S_{\mathrm{gap}}^2 =0.05$ ($S_{\mathrm{gap}}^2=0.03$) 
at the Tevatron (LHC). Within the same model, we also analyze the 
Pomeron--photon contribution in a way analogous to the Pomeron--odderon 
contribution. The values of the phenomenologically improved 
cross-sections are summarized in Table~1\footnote{
The photon cross-sections given in Table~1 are not meant to provide the 
most accurate estimate, but rather to asses the impact of model 
assumptions on the odderon/photon ratio.}. 
The photon and the odderon contributions do not interfere 
in the lowest order approximation and the corresponding cross-sections 
may be treated independently.
As seen from the table, the Pomeron--odderon contributions are found to be 
uncertain, with a multiplicative uncertainty factor of 3--5. The ambiguities, 
however, cancel partially in the ratio of the Pomeron--odderon contribution 
to the Pomeron--photon contribution evaluated in the same scenario. 
Thus, within the considered scenarios, the ``odderon to photon ratio'' 
$R=[d\sigma^{\mathrm{corr}} /dy] / [d\sigma^{\mathrm{corr}} _{\gamma}/dy]$
varies between 0.3 and~0.6 for $J/\psi$ production at the Tevatron,
and between about 0.06 and~0.15 at the LHC. 
In the case of $\Upsilon$, $R$~varies between about 0.8 and 1.7 at 
the Tevatron and  between about 0.15 and 0.4 at the LHC. 
These numbers suggest that the odderon contribution may well 
be of a similar magnitude to the photon contribution at the Tevatron 
and somewhat smaller than the photon contribution at the LHC.

\noindent
\begin{figure}
\begin{tabular}{ll}
\epsfxsize=0.45\columnwidth 
\epsffile{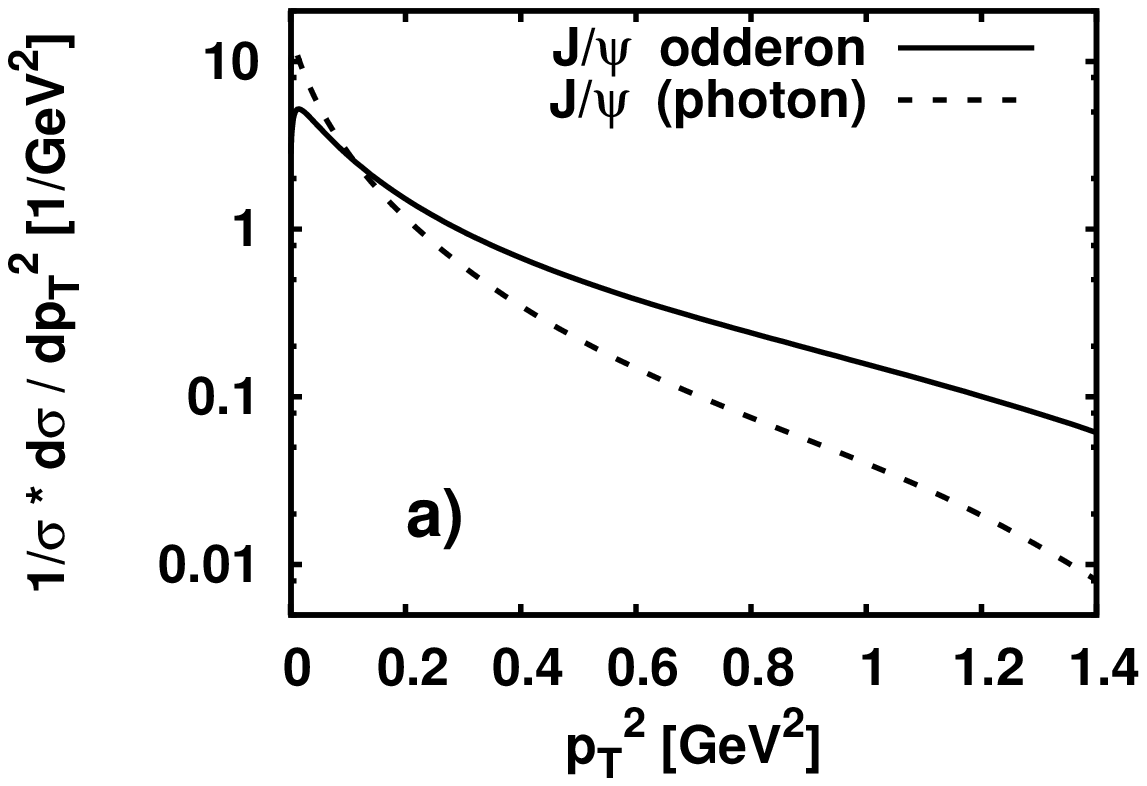} & 
\epsfxsize=0.45\columnwidth
\epsffile{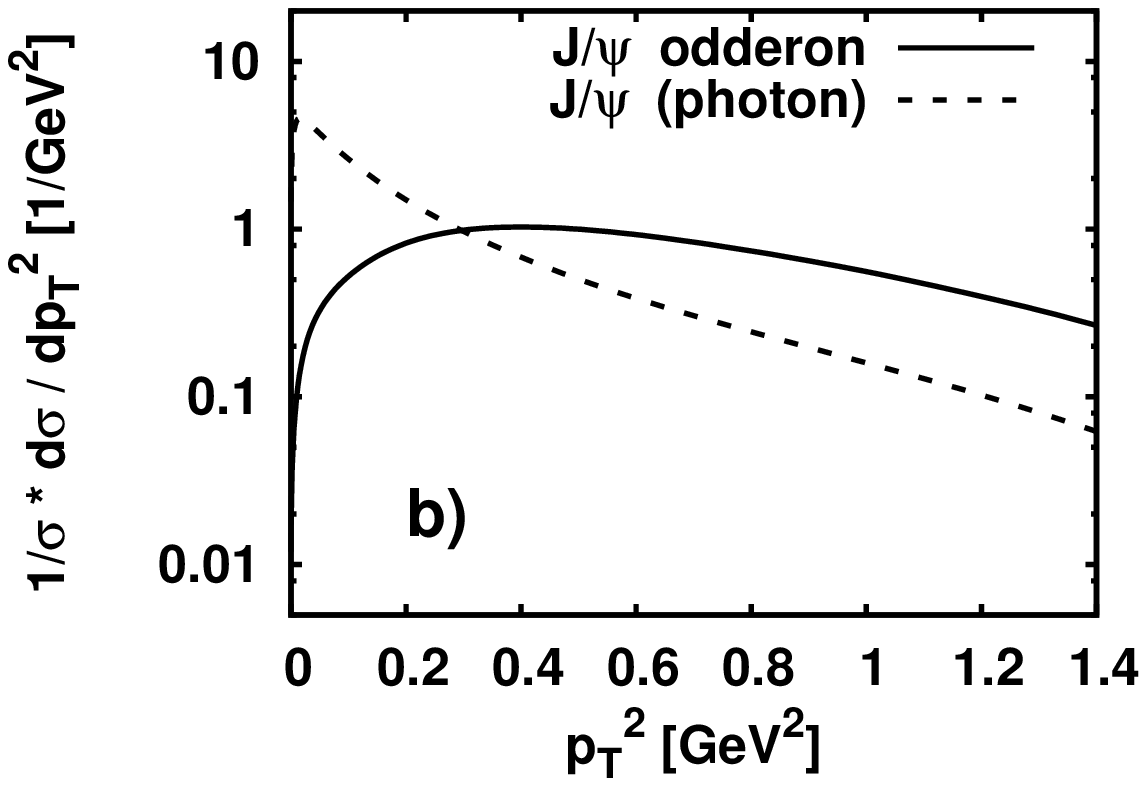}
\end{tabular}
\caption{The normalized differential cross-section 
$\,(d\sigma/dp_T^2)/\sigma\,$ for 
the exclusive production: a)~$\,p\bar p \to p \bar p \,J/\psi\,$,
and b)~$\,p p \to p p \,J/\psi\,$ as a function
of meson $p_T^2$ for the odderon--Pomeron and the photon--Pomeron fusion. 
}
\end{figure}

The obtained results clearly indicate that the photon exchange background 
to the odderon mediated processes is important in the $p_T$-integrated
cross-sections at central rapidities. Thus in the search for the 
odderon one should use also the available information on the transverse
momentum distributions. In Fig.~2 normalized distributions
of the meson~$p_T ^2$ are shown for the odderon and photon contributions to
the exclusive $J/\psi$ production (the results for $\Upsilon$ are similar).
As seen from this figure, the relative importance of the odderon 
contribution increases at larger meson $p_T$, and the different $p_T$-shape
of the odderon and the photon contributions may be used to perform
an experimental separation between them. Another measurement 
with an enhanced sensitivity to the odderon should be possible
using forward proton detectors. Assuming, that a forward 
detector at the LHC measures the proton~$A$ that lost about 
$x_A \sim 0.01$ of its energy in the $pp \to pp \Upsilon$ process, 
one finds that the other proton, $B$, should lose a tiny fraction of 
about $x_B \sim 10^{-4}$ of its energy. Although proton~B cannot be 
measured, the asymmetric kinematics implies that the proton~$B$ couples 
predominantly to the Pomeron and proton~$A$ couples to the photon or the 
odderon. The photon exchange is characterized by a steep $1/p_T^2$ behavior 
and it leads to the $p_T$~distribution of proton~$A$ that is concentrated
at small momenta.  The odderon induced $p_T$-distribution is much 
broader. Thus, for instance, a cut of $p_T > 0.5$~GeV on the transverse
momentum of proton~A  
should increase the odderon to photon ratio in the data sample by a 
factor of about~10.

In summary, measurements of the exclusive heavy vector meson production
at the Tevatron and the LHC may provide a viable opportunity to discover 
the  odderon.


\begin{footnotesize}


\end{footnotesize}


\end{document}